\documentclass[conference]{IEEEtran}
\IEEEoverridecommandlockouts
\usepackage{cite}
\usepackage{url}
\usepackage{booktabs}
\usepackage{amsmath,amssymb,amsfonts}
\PassOptionsToPackage{hyphens}{url}\usepackage{hyperref}
\usepackage[capitalise]{cleveref}
\usepackage{algorithmic}
\usepackage{graphicx}
\usepackage{textcomp}
\let\oldFootnote\footnote
\newcommand\nextToken\relax

\renewcommand\footnote[1]{%
    \oldFootnote{#1}\futurelet\nextToken\isFootnote}

\newcommand\isFootnote{%
    \ifx\footnote\nextToken\textsuperscript{,}\fi}

\usepackage[table]{xcolor}

\usepackage{subcaption}
\usepackage[colorinlistoftodos]{todonotes}

\def\BibTeX{{\rm B\kern-.05em{\sc i\kern-.025em b}\kern-.08em
    T\kern-.1667em\lower.7ex\hbox{E}\kern-.125emX}}
\begin{document}

\title{MLAAD: The Multi-Language Audio Anti-Spoofing Dataset\\
}

\makeatletter
\newcommand{\linebreakand}{%
  \end{@IEEEauthorhalign}
  \hfill\mbox{}\par
  \mbox{}\hfill\begin{@IEEEauthorhalign}
}
\makeatother

\author{
\makebox[\textwidth][c]{%
Nicolas M. M\"uller\textsuperscript{1,2},
Piotr Kawa\textsuperscript{3,2},
Wei Herng Choong\textsuperscript{1},
Edresson Casanova\textsuperscript{4},}\\
\makebox[\textwidth][c]{%
Eren Gölge\textsuperscript{4},
Thorsten M\"uller\textsuperscript{5},
Piotr Syga\textsuperscript{3},
Philip Sperl\textsuperscript{1},}\\
\makebox[\textwidth][c]{%
Konstantin Böttinger\textsuperscript{1}}\\[1ex]
\makebox[\textwidth][c]{%
\textsuperscript{1}Fraunhofer AISEC, Germany\quad
\textsuperscript{2}Resemble AI, USA\quad
\textsuperscript{3}Wroc{\l}aw University of Science and Technology, Poland}\\
\makebox[\textwidth][c]{%
\textsuperscript{4}Coqui.ai, Germany\quad
\textsuperscript{5}Thorsten-Voice}\\[1ex]
\makebox[\textwidth][c]{\texttt{nicolas.mueller@aisec.fraunhofer.de}}
}

\maketitle

\begin{abstract}
This paper presents the Multi-Language Audio Anti-Spoofing Dataset (MLAAD), version 11: a dataset of synthetic audio to train and evaluate audio deepfake detection models.
It features 205Text-to-Speech (TTS) models, comprising a total of 1153.6hours of synthetic voice in 54different languages.

To evaluate this dataset, we train three state-of-the-art deepfake detection models with MLAAD and observe that it demonstrates superior performance to comparable datasets like InTheWild and FakeOrReal when used as a training resource. Moreover, compared to the renowned ASVspoof 2019 dataset, MLAAD proves to be a complementary resource. In tests across eight datasets, MLAAD and ASVspoof 2019 alternately outperformed each other, each excelling on four datasets.

By publishing the dataset\footnote{\url{https://deepfake-total.com/mlaad}}
and making a trained model accessible via an interactive webserver\footnote{\url{https://deepfake-total.com/}}
, we aim to democratize anti-spoofing technology, making it accessible beyond the realm of specialists, and contributing to global efforts against audio spoofing and deepfakes.
\end{abstract}


\section{Introduction}
In recent years, text-to-speech (TTS) technology has significantly advanced due to developments in machine learning (ML), leading to a wide range of beneficial applications~\cite{parrotron,ios-personal-voice}. This includes providing speech assistance to individuals with disabilities or health-related speech impairments. However, the evolution of TTS also brings notable challenges, particularly in the form of deepfakes and audio spoofs. The former are sophisticated fakes that pose risks by deceiving humans, potentially leading to fraud, misinformation, and the spread of fake news~\cite{audio-scam}. 
The latter can be used to compromise biometric identification systems~\cite{asvspoof2019}.

Addressing these concerns requires a multifaceted approach.
First and foremost, this includes public education on media literacy. Additionally, the Content Authenticity Initiative~\cite{content-auth-init}, spearheaded by Adobe, is focused on countering misinformation by employing cryptographic methods to validate images and audio. A critical component is the advancement of reliable detection techniques for audio spoofs and deepfakes, which remains an active field of research.

Despite these efforts, detecting audio deepfakes and spoofs remains a challenging task. This difficulty is partly due to a discrepancy between laboratory conditions and real-world environments; models that perform well in controlled settings often fail in practical applications~\cite{in-the-wild}. Another significant challenge is the language bias in deepfake audio data, which is predominantly in English, thus limiting its applicability and potentially discriminating against non-English speakers.
With the increasing prevalence of suspected deepfakes in various non-English media~\cite{aiputin,df_france}, the need for a multilingual audio deepfake detection system is imperative.

\begin{table}[]
    \centering
\resizebox{.49\textwidth}{!}{
\begin{tabular}{@{}lccccc@{}}
\toprule
 & Release Date       & Hours & Models & Languages & Distributed via    \\ \midrule
v1 & 17.01.2024 & 63.0 & 21 & 22 & zip              \\
v2 & 28.02.2024 & 63.0 & 21 & 22 & zip              \\
v3 & 16.04.2024 & 163.9 & 54 & 23 & zip               \\
v4 & 24.09.2024 & 175.0 & 59 & 23 & zip               \\
v5 & 02.11.2024 & 398.3 & 87 & 38 & zip + Huggingface \\
v6 & 26.04.2025 & 420.7 & 91 & 38 & Huggingface       \\
v7 & 11.07.2025 & 485.3 & 101 & 40 & Huggingface     \\ 
v8 & 01.10.2025 & 570.3 & 119 & 40 & Huggingface     \\ 
v9 & 17.01.2026 & 687.4 & 140 & 51 & Huggingface     \\
v10 & 18.05.2026 & 1002.9 & 175 & 54 & Huggingface     \\
v11 & 01.09.2026 & 1153.6 & 205 & 54 & Huggingface     \\ \bottomrule

\end{tabular}

}
    \caption{Overview of MLAAD dataset versions, including release dates, total duration of synthesized speech, number of TTS models, supported languages, and distribution method.}
    \label{tab:mlaad_versions}
\end{table}
To address these issues, we introduce and continuously expand our MLAAD dataset (c.f.~\Cref{tab:mlaad_versions}), a large-scale corpus of state-of-the-art audio fakes encompassing 54languages. 
This dataset, based on the M-AILABS Speech Dataset~\cite{mai},
includes 1153.6hours of synthesized speech, created by 205state-of-the-art TTS models. 
By training state-of-the-art deepfake detection models using MLAAD, we show that the resulting models achieve high performance scores in cross-dataset evaluations, indicating practicality in real-world conditions.

We open-source the MLAAD dataset and make our models interactively accessible to a broader audience, including non-technical individuals, through the platform \url{https://deepfake-total.com}.

\section{Related Work}

\begin{table}[]
    \centering
    \begin{tabular}{llll}
    \toprule
    name & languages & \# systems & \# utterances \\
    \midrule
    ASVspoof15~\cite{asvspoof2015}    & English & 10 & 263,151 \\
    ASVspoof19 LA~\cite{asvspoof2019} & English & 19 & 121,461\\
    ASVspoof21 LA~\cite{asvspoof2021} & English & 13 & 164,612  \\
    ASVspoof21 DF~\cite{asvspoof2021} & English & 100+ & 593,253 \\
    
    FakeAVCeleb~\cite{fakeavceleb}     & English & 1 & 11,857 \\
    
    FoR~\cite{for}             & English & 7 
    & 195,541 \\ 
    Voc.v~\cite{vocoders-nii} & English & 8 & 82,048 \\
    In-The-Wild~\cite{in-the-wild}     & English & ? & 31,779 \\
    PartialSpoof~\cite{partial-spoof} & English & 19 & 121,461 \\ 
    WaveFake~\cite{wavefake}        & English, Japanese & 9 & 136,085 \\
    ADD 2022~\cite{add2022}        & Chinese & ? & 493,123 \\ 
    ADD 2023~\cite{add2023}        & Chinese & ? & 517,068 \\ 
    FMFCC-A~\cite{fmfcca}         & Chinese & 13 & 50,000 \\
    HAD~\cite{had}             & Chinese & 2 & 160,836 \\
    CFAD~\cite{cfad} & Chinese & 12 & 347,400 \\
    
    \textbf{MLAAD (ours)}  & \textbf{54} 
    & \textbf{205} & \textbf{534,000} \\
    \bottomrule
\end{tabular}
    \caption{Overview of Related Work on Datasets Used in Audio Anti-spoofing Research.}
    \label{tab:related_work_audio}
\end{table}

\subsection{Speech Synthesis}
In the realm of speech synthesis, we primarily identify two main approaches: 
Text-to-speech (TTS) and Voice Conversion (VC). 
Text-to-speech encompasses algorithms that generate speech from text, where the voice, accents, and pitch characteristics are derived from the training dataset~\cite{tacotron2,speedy-speech}. 
Modern TTS methods have evolved to allow a single model to produce multiple voices~\cite{tortoise-tts,xtts,vits,capacitron,your-tts,glow-tts} and languages~\cite{xtts,your-tts}. 
In contrast, Voice Conversion involves creating speech from two inputs – one for linguistic information (text) and another for vocal characteristics (voice), resulting in an utterance that combines the text of one input with the voice of another~\cite{xtts,vits,stargan-vc,freevc}.
The field is supported by numerous open-source toolkits~\cite{espnet,coqui-ai,speechbrain, huggingface}, which offer access to various speech synthesis methods.
Especially Coqui-TTS~\cite{coqui-ai} and Hugging Face~\cite{huggingface} encompass many state-of-the-art approaches. Different datasets, varying in language, number of speakers, and accents, are used to train these methods~\cite{ljspeech,blizzard_2013,css_10,common-voice}.

\subsection{Learning Models for Deepfake Detection}
The increasing advancement and popularity of speech synthesis techniques, however, have also enabled misuse, disinformation, and fraud. 
This has prompted research in the domain of neural voice anti-spoofing, which aims to classify a given audio file as authentic ('bona-fide') or fake ('spoof').
Most current methods rely on deep neural networks, falling into two main categories based on the type of information processed. 
First, raw audio signal-based models work directly with the raw waveform, eliminating the need for transforming the input, and include methods like~\cite{rawnet2,rawgat-st,aasist,raw-pcdarts}. 
Second, front-end-based models analyze transformed representations of audio signals, such as (mel-)spectrograms or cepstral coefficients~\cite{sahidullah15_interspeech,lcnn_nii}, and have recently been enhanced by Self-Supervised Learning or embedding-based approaches due to their high performance and good generalization capabilities~\cite{lcnn_nii,mesonet,w2v-df,w2v-vicomtech,whisper-df}. 
Additionally, the development of complex-valued networks for voice anti-spoofing~\cite{muller2023complex} is an emerging area of interest in this field.

\subsection{Datasets for Deepfake Detection}
Several voice anti-spoofing datasets 
have been released for training these models, c.f.~\cref{tab:related_work_audio}.
These datasets focus on enhancing weak model generalization capability by including a wide array of generation methods, codecs, noises, and data quality~\cite{asvspoof2021,wavefake,add2022,add2023,cfad}, as well as dealing with partial spoofs~\cite{partial-spoof,had,add2022}, real-world deepfake instances~\cite{in-the-wild}, and multimodal deepfakes combining fake audio and video~\cite{fakeavceleb,av-deepfake1m}. 
However, a gap remains in addressing the diversity of languages in spoofed utterances, with most datasets focusing predominantly on English~\cite{asvspoof2021,partial-spoof,in-the-wild,wavefake,fakeavceleb} and Chinese~\cite{had,cfad,fmfcca,add2022,add2023}. 
Although WaveFake~\cite{wavefake} incorporates both English and Japanese, datasets predominantly featuring English continue to dominate the field of anti-spoofing.
To the best of our knowledge, there is no speech anti-spoofing dataset that comprehensively addresses multiple languages.

\section{MLAAD -- Dataset Description}

\begin{figure}
    \centering
    \includegraphics[width=0.49\textwidth]{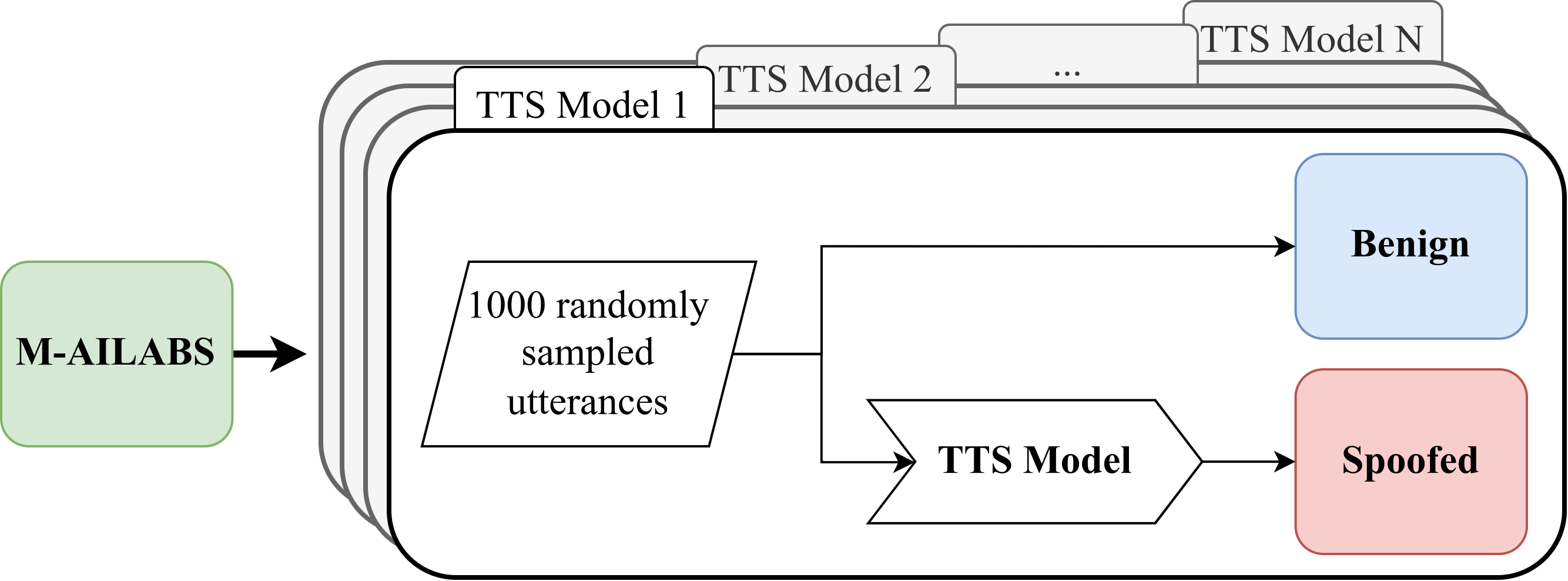}
    \caption{Visualization of the data creation process for MLAAD.}
    \label{fig:creation_of_mlaad}
\end{figure}

\subsection{Synthesis Procedure}
We introduce the MLAAD dataset, an extension of the M-AILABS Speech Dataset, which features audio clips in eight original languages: English, French, German, Italian, Polish, Russian, Spanish, and Ukrainian.
All of these audio clips represent authentic human speech, taken from audiobooks or speeches and interviews of public figures, such as former German Chancellor Angela Merkel.
We enrich this dataset with computer-generated audio clips in 54languages.

Our dataset construction involves the following steps, as illustrated in \cref{fig:creation_of_mlaad}:
First, for each combination of language and TTS model, we select a random subset of 1000 instances from the original dataset. 
If the target language is present in M-AILABS, we use these samples as a starting point.
Otherwise, we use the texts of the English samples and translate them to the target language using Neural Machine Translation~\cite{nmt_pypi}.

Second, for each of the baseline audio files, a corresponding synthetic version is created.
The text for this synthesis is sourced from the original audio files or the translated version thereof. 
Additionally, we use a speaker reference file for models supporting multi-speaker input, such as VITS trained on multi-speaker datasets.
This speaker reference file is chosen randomly from the same set of baseline audio files. 
For models trained on single-speaker datasets like LJSpeech~\cite{ljspeech}, such a speaker reference is not utilized. 

Third, we aggregate the results: information on the synthesized audio file, its language, and the transcript are stored in a \textit{meta.csv} file within the corresponding folder. 
These files constitute the ``fake'' audio files when used for supervised learning, while the original baseline audio files are used as ``bona-fide'' training samples.
We use 22,050 Hz, with a 16-bit depth, and store the audio in WAV format.
The TTS models are derived from Coqui.ai and Hugging Face and encompass nearly all current state-of-the-art TTS models, c.f.~\cref{tab:arch}.
Additionally, for each of the original languages present in M-AILABS, we use Griffin Lim~\cite{griffin1984signal} as a model drop-in to re-synthesize 1000 randomly chosen audio files.

\subsection{Formatting}
\begin{table}
\centering
\begin{tabular}{|l|}
\hline
        Descriptor \\
\hline
                path \\
       original\_file \\
            language \\
is\_original\_language \\
            duration \\
       training\_data \\
          model\_name \\
        architecture \\
          transcript \\
          reference\_speaker (since  \textit{v8}) \\
\hline
\end{tabular}
\caption{Descriptors in the MLAAD dataset.}
\label{tab:datafields}
\end{table}

The outputs are organized into directories describing language and model type.
Each directory contains the synthesized audio files and an accompanying ``meta.csv'' file, which encapsulates various data fields as indicated in ~\cref{tab:datafields}. 
The ``path'' field specifies the location of the synthesized audio file, relative to the base path of MLAAD. The ``original\_file'' field denotes the filepath of the baseline audio, relative to the base path of the M-AILABS dataset. The ``language'' field identifies the language of the audio file, while ``is\_original\_language'' indicates if the language was originally included in M-AILABS, or if the transcript has been translated. 
The ``duration'' field shows the length of the audio file in seconds, and ``training\_data'' reveals the dataset used to train the TTS model. The ``model\_name'' field provides a unique identifier for the model, such as ``facebook/mms-tts-eng'' for the Facebook Massively Multilingual Speech (MMS) TTS model~\cite{mms_tts}, while ``architecture'' describes the TTS model architecture (such as VITS), and   ``transcript'' represents the textual transcript of the audio file.
Lastly, ``reference\_speaker'' specifies the reference speaker used to generate the audio file.
Depending on the TTS model, this value is either a file path pointing to an M-AILABS audio file or a speaker name (for models that provide a fixed set of speakers).
Note that this key was introduced only in Version ``v8'', and may therefore be missing from previous ``meta.csv'' files.

\subsection{Statistics}

\begin{table*}[p]
    \centering
    \small
    \begin{tabular*}{\textwidth}{@{\extracolsep{\fill}}l r | l r | l r@{}}
\toprule
\textbf{Architecture} & \textbf{Hours} & \textbf{Architecture} & \textbf{Hours} & \textbf{Architecture} & \textbf{Hours} \\
\midrule
Azzurra-Voice & 1.5 & Llasa & 30.9 & Sesame CSM-1B & 1.7 \\
Bark & 123.3 & LongCat-AudioDiT & 3.3 & Soprano11 & 1.6 \\
Capacitron & 1.5 & LuxTTS & 1.8 & SopranoTTS & 1.7 \\
CaroTTS & 1.7 & MOSS-TTS & 41.0 & Sopro & 2.0 \\
Cartesia.ai (Sonic-3) & 3.5 & Mars5 & 1.8 & SoulX-Podcast & 1.9 \\
ChatTTS & 3.7 & MatchaTTS & 1.7 & Spark TTS & 2.1 \\
Chatterbox & 16.3 & Maya1 TTS & 2.1 & SpeechT5 & 2.0 \\
Coda & 7.3 & MegaTTS3 & 4.1 & Speedy Speech & 2.0 \\
DeepGram & 3.8 & MeloTTS & 1.8 & Step-Audio-EditX & 3.8 \\
DiT & 2.0 & Metavoice-1B & 1.9 & StyleTTS 2 & 4.1 \\
Dramabox & 3.5 & Microsoft VibeVoice & 8.3 & StyleTTS2 & 2.1 \\
Dual-AR & 19.8 & Ming-omni-tts & 3.9 & Supertonic & 17.7 \\
DualAR & 3.6 & MiniCPM-o-2.6 & 4.4 & SvaraTTS v1 & 5.2 \\
E2 TTS & 1.8 & MiniMax-Speech & 2.0 & TADA & 17.5 \\
Edge-TTS & 39.9 & MiraTTS & 2.0 & Tacotron & 2.5 \\
ElevenLabs & 10.1 & MisoTTS & 2.3 & Tacotron 2 & 20.6 \\
F5 TTS & 1.7 & NVidia Magpie-TTS & 4.5 & Tortoise & 2.1 \\
F5-TTS & 18.5 & Nari Dia-1.6B & 1.9 & VITS & 58.1 \\
FastPitch & 2.1 & Nari Dia2 & 2.1 & VITS Neon & 4.3 \\
FireRedTTS & 5.5 & NeuTTS & 2.0 & VITS-MMS & 23.3 \\
FishTTS & 7.4 & NeuTTS-Air & 2.0 & VITS2 & 1.7 \\
FreyaTTS & 1.6 & NeuralHMM & 2.1 & Veena & 3.5 \\
GLM-TTS & 2.1 & OmniVoice & 53.6 & VibeVoice & 2.0 \\
GPA & 3.5 & OpenAI TTS-1 HD & 13.7 & VieNeu-TTS & 2.3 \\
GPT-SoVITS & 2.1 & OpenVoiceV2 & 4.3 & VixTTS & 3.6 \\
Gemini & 2.5 & Openaudio-S1-Mini & 4.8 & VoXtream & 1.9 \\
GlowTTS & 12.6 & Optispeech & 1.5 & VoiceCore & 2.9 \\
Griffin Lim & 16.6 & Orpheus TTS v0.1 & 2.0 & VoxCPM & 35.6 \\
Higgs-Audio & 15.5 & OuteTTS & 15.1 & VoxCPM-1.5 & 4.1 \\
Higgs-Audio-V2 & 8.8 & Overflow & 2.2 & Voxtral & 19.1 \\
Index-TTS & 7.6 & Parler TTS & 4.9 & Voxtream & 2.2 \\
IndicF5 & 16.6 & Parler-TTS & 10.0 & WavTTS & 1.7 \\
Indri-TTS-0.1 & 3.7 & PocketTTS & 1.9 & WhisperSpeech & 3.9 \\
Inflect-Nano & 1.5 & PrimeTTS & 2.1 & XTTS v1.1 & 36.9 \\
Inworld-TTS & 3.9 & Qwen2.5-Omni & 2.3 & XTTS v2 & 38.9 \\
Jenny & 1.7 & Qwen3-Omni & 34.3 & YarnGPT & 5.2 \\
Kani-TTS & 3.8 & Qwen3-TTS & 2.3 & ZipVoice & 3.4 \\
Kitten-TTS & 4.1 & RVC & 17.6 & Zonos & 8.8 \\
Kokoro & 10.4 & Raon-OpenTTS & 1.8 & ZonosTTS v0.1 & 3.7 \\
KugelAudio & 15.6 & Resemble.ai (April 12th, 2025) & 8.8 & dots.tts & 45.5 \\
Kyutai-TTS & 4.5 & Ringg Squirrel TTS & 3.9 & minimax/speech-02 & 8.8 \\
LFM2.5-Audio & 1.8 & Sarashina & 2.2 &  &  \\
Lightning & 3.5 & Sesame CSM 1B & 2.2 &  &  \\
\bottomrule
\end{tabular*}

    \caption{TTS model architectures utilized in MLAAD along with the corresponding duration of the synthesized audio files (in hours).}
    \label{tab:arch}
\end{table*}

\begin{table*}[p]
    \centering
    \small
    \begin{tabular*}{\textwidth}{@{\extracolsep{\fill}}l r | l r | l r@{}}
\toprule
\textbf{Language} & \textbf{Hours} & \textbf{Language} & \textbf{Hours} & \textbf{Language} & \textbf{Hours} \\
\midrule
Amharic & 3.6 & Hungarian & 12.5 & Polish & 44.5 \\
Arabic & 19.4 & Igbo & 1.8 & Portuguese & 39.2 \\
Bangla & 14.1 & Indonesian & 4.5 & Romanian & 12.8 \\
Bulgarian & 6.1 & Irish & 6.4 & Russian & 50.2 \\
Chinese & 54.6 & Italian & 57.3 & Sinhala & 2.1 \\
Croatian & 1.8 & Japanese & 29.9 & Slovak & 6.3 \\
Czech & 13.0 & Javanese & 2.4 & Slovenian & 1.8 \\
Danish & 7.7 & Kannada & 9.4 & Spanish & 66.6 \\
Dutch & 20.8 & Korean & 30.2 & Swahili & 3.4 \\
English & 286.5 & Latvian & 2.3 & Swedish & 7.4 \\
Estonian & 5.7 & Lithuanian & 2.3 & Tamil & 5.0 \\
Finnish & 13.2 & Luxembourgish & 1.7 & Thai & 7.8 \\
French & 84.6 & Malay & 2.5 & Turkish & 15.6 \\
German & 84.6 & Malayalam & 1.8 & Turkmen & 3.9 \\
Greek & 8.8 & Maltese & 8.1 & Ukrainian & 35.7 \\
Hausa & 1.5 & Marathi & 3.0 & Urdu & 2.1 \\
Hebrew & 2.0 & Norwegian & 2.0 & Vietnamese & 10.1 \\
Hindi & 22.5 & Persian & 8.8 & Yoruba & 1.8 \\
\bottomrule
\end{tabular*}

    \caption{Languages utilized in MLAAD along with the corresponding duration of the synthesized audio files (in hours).}
    \label{tab:lang}
\end{table*}

The dataset encompasses 1153.6hours of synthesized speech across 54languages, generated using 205state-of-the-art TTS models, comprising 127different architectures.
We distribute only these synthesized samples, for the originals please refer to the M-AILABS dataset.

\Cref{tab:arch} and \cref{tab:lang} present the architecture and language distribution within MLAAD, offering an overview of the TTS architectures and languages that constitute the dataset.







\section{Evaluation}

\begin{table*}[ht!]
\centering
\resizebox{.999\textwidth}{!}{
\begin{tabular}{l|lllllllll}
\toprule
 &             & Test Data $\rightarrow$ & &&&&&&\\
Model &      Train Data $\downarrow$        & ASVspoof19 & ASVspoof21-DF & ASVspoof21-LA & FakeOrReal & InTheWild & MLAAD v1 & Voc.v & WaveFake \\
\midrule
RawGat-ST & ASV19 &                           - &                            67.1±10.4 &   -                          &                       68.8±11.2 &                        50.0±2.5 &                          56.9±5.7 &                         37.6±23.7 &                            15.0±16.0 \\
 &   FOR &                         49.1±18.1 &                             53.6±3.8 &                             56.4±6.9 &                         - &                        49.8±0.4 &                          51.9±3.3 &                         25.4±43.1 &                            20.0±44.7 \\
 &   ITW &                         58.4±10.2 &                             58.5±4.2 &                             55.1±2.9 &                        54.5±3.9 &                         - &                          57.4±7.0 &                         38.9±27.9 &                            65.3±30.3 \\
 & MLAAD v1 &                         60.9±17.8 &                             50.4±1.1 &                             50.5±2.1 &                        50.2±0.4 &                        47.7±3.3 &                           - &                         70.5±42.2 &                            68.4±41.7 \\ \midrule
                        SSL W2V2 & ASV19 &                           - &                  \cellcolor{blue!15}           91.6±3.8 &                           -   &             \cellcolor{blue!15}            81.1±7.7 &                      \cellcolor{blue!15}   79.7±6.8 &                \cellcolor{blue!15}           71.8±5.1 &                         71.6±12.1 &                            51.3±28.5 \\
                         &   FOR &                         65.4±10.3 &                             77.7±2.6 &                     \cellcolor{blue!15}         73.5±3.5 &                         - &                       57.8±10.9 &                          57.1±3.4 &                         21.5±18.4 &                              1.5±2.0 \\
                         &   ITW &                         65.0±10.1 &                             68.1±4.9 &                             60.2±4.7 &                        55.3±5.7 &                         - &                          59.1±4.3 &                         19.2±17.1 &                            70.4±35.4 \\
                         & MLAAD v1 &             \cellcolor{blue!15}            78.0±15.3 &                            75.5±15.7 &               \cellcolor{blue!15}              73.7±10.4 &                        64.4±9.0 &                       68.0±17.5 &                           - &                         66.0±34.2 &                            69.8±38.4 \\  \midrule
                       Whisper DF & ASV19 &                           - &                             82.3±3.8 &                           -   &                        80.6±4.4 &                        76.5±4.0 &                          44.9±3.3 &                         18.1±14.2 &                              2.2±3.5 \\
                       &   FOR &                          45.9±0.8 &                             60.3±1.4 &                             61.1±2.3 &                         - &                        54.1±4.3 &                          54.5±1.1 &                           1.3±0.9 &                              0.2±0.1 \\
                       &   ITW &                          55.5±9.3 &                             63.8±3.6 &                             61.7±4.4 &                        67.2±5.6 &                         - &                          54.2±3.5 &                         18.4±19.2 &                            26.3±40.0 \\
                       & MLAAD v1 &                          70.8±0.9 &                             52.7±6.1 &                             52.6±6.2 &                        50.5±3.3 &                        54.3±4.9 &                           - &         \cellcolor{blue!15}                 86.7±29.4 &              \cellcolor{blue!15}                97.2±3.3 \\
\bottomrule
\end{tabular}
}
\caption{Summary of training results across different datasets, measured by test accuracy. Each row represents a model trained on one dataset and evaluated on the remaining seven datasets, as described in \cref{sec:eval1}.
Highlights indicate the highest value per test dataset (i.e., per column).
Averages are calculated from five separate trials, presented as mean ± standard deviation.}
\label{tab:exp1}
\end{table*}

\subsection{Training Setup}
To evaluate our dataset, we selected three state-of-the-art voice anti-spoofing models: 
RawGat-ST~\cite{rawgat-st}, SSL-W2V2~\cite{ssl_antispoof}, and WhisperDF~\cite{whisper-df}.

The RawGat-ST and SSL-W2V2 models process 5-second clips of raw audio, whereas WhisperDF handles 30-second audio segments, converting them into a time-frequency representation through its frontend. 
We augment all training data with random noise sourced from the RIRS Noises~\cite{rirs-db} and Noise ESC-50~\cite{esc-50} datasets, along with music randomly selected from the Free Music Archive (Instrumental Music)~\cite{free-music-archive} and Musan~\cite{musan}, overlaying randomly chosen noise or music with a 5\% probability per sample. 
Additionally, with a 20\% probability, each sample is encoded using one of several codecs: ulaw, alaw, mp3, aac, flac, opus, ac3.
These augmentations serve to diversify the dataset and exclude learning shortcuts~\cite{muller2021speech}, such as the model learning to associate a certain encoding with either fake or real data.

To ensure balanced training, the data is randomly undersampled for each training epoch.
This exposes the models to an equal number of fake and authentic audio samples, preventing majority bias. 
All three models are trained for 100 epochs, with the provision for early stopping based on training accuracy.

Contrary to typical voice anti-spoofing evaluations \cite{ssl_antispoof, muller2023complex, asvspoof2019, asvspoof2021} that use the equal error rate (EER), we opt for accuracy as our evaluation metric. 
This decision is made because EER can only be calculated when test data contains both authentic and spoof samples. 
However, some of our test datasets, like Voc.v~\cite{vocoders-nii}, consist only of spoof samples. 
We believe that the accuracy metric better suits our real-world focus since it measures actionable output, unlike EER, which relies on an optimal threshold that is unknown to practitioners without ground-truth data.

\subsection{Cross-Dataset Evaluation}\label{sec:eval1}
These learning models are trained separately on four different datasets: MLAAD v1, ASVspoof19~\cite{asvspoof2019}, FakeOrReal (FoR)~\cite{for}, and In-The-Wild (ITW)~\cite{in-the-wild}. 
Evaluation is conducted on eight datasets, including the four used for training, as well as ASVspoof21-DF and ASVspoof21-LA~\cite{asvspoof2021}, Voc.v~\cite{vocoders-nii}, and WaveFake~\cite{wavefake}. 
We do not use the latter four datasets for training because ASVspoof21-LA and DF are intended solely as test-sets~\cite{asvspoof2021}.
Voc.v and WaveFake consist only of fake samples, precluding supervised training.
Finally, we refrain from using ASVspoof21-LA to assess the cross-dataset performance of models trained on ASVspoof19, given the substantial similarity between the two datasets.
Note that the results reported here use MLAAD v1; the evaluation was not re-run for subsequent releases, but we expect later versions to perform substantially better given the much larger volume and diversity of training data.

For every configuration, we conduct the experiment five times, reporting the mean accuracy along with the standard deviation in \cref{tab:exp1}. 
The results quantify a dataset's ability to enable state-of-the-art models to generalize voice anti-spoofing to cross-dataset data, a key requirement for real-world applicability.

We observe the following:
First, no single training dataset consistently outperforms the others. 
Notably, ASVspoof19 and MLAAD each achieve the highest accuracy in four out of eight cross-dataset test cases. 
In contrast, datasets like FakeOrReal and InTheWild do not top any case. 
Second, the shared success of ASVspoof19 and MLAAD suggests that they complement each other in terms of their training effectiveness. 

Third, we observe that the models sometimes seem to learn features that, paradoxically, seem to transfer very well to some datasets, while being completely useless for others.
Take for example the SSL W2V2 model, trained on ASVspoof19.
It shows excellent performance on ASVspoof21-DF and good performance on FakeOrReal and InTheWild, while at the same time failing completely on WaveFake, where it achieves 51\% accuracy, which is as good as random guessing.
For some test cases, the models' accuracy scores are even lower than 50\%, indicating that the training data comprises shortcuts that not only are not helpful but actually detrimental to the performance.
Notably, models trained on MLAAD are the only ones that do not exhibit performance significantly below 50\%, hinting at possibly less detrimental shortcut features comprised in MLAAD than in other datasets~\cite{muller2021speech}.

\section{Conclusion} 
In this paper, we present a new audio dataset, which comprises voice spoofs in 54languages by 205TTS models, composed of 127architectures.
When evaluating MLAAD's usefulness as a training resource, we observe that MLAAD complements existing research.
It consistently outperforms related work such as InTheWild or FakeOrReal with respect to the trained models' cross-dataset generalization capability.
Furthermore, in comparison with the renowned ASVspoof 2019 dataset, MLAAD proves to be a complementary resource. Across eight datasets, MLAAD and ASVspoof 2019 alternately outperformed each other, both excelling on four datasets:
Training the SSL W2V2 model on ASVspoof19 results in the highest performance on ASVspoof21-DF, FakeOrReal, InTheWild, and MLAAD. Similarly, training SSL W2V2 or WhisperDF on MLAAD leads to optimal performance on ASVspoof19, ASVspoof21-LA, Voc.v, and WaveFake.
We note that an evaluation of the multi-lingual anti-spoofing capabilities has to remain a topic for future work since there are no other comparable multi-lingual datasets against which to evaluate MLAAD.
However, the strong results on WaveFake, which comprises both English and Japanese, suggest that the learning models profit from the diversity of languages in MLAAD.

\section{Acknowledgment} 
This work has been (partially) funded by the Bavarian Ministry of Economic Affairs, Regional Development and Energy; and the Department of Artificial Intelligence, Wrocław University of Science and Technology.



\bibliographystyle{IEEEtran}
\bibliography{bibliography}



\end{document}